\documentclass{revtex4-2}

\usepackage{graphicx}
\usepackage{dcolumn}
\usepackage{bm}
\usepackage{chemformula}
\usepackage{xcolor}

\renewcommand{\figurename}{Fig.}

\begin{document}

\preprint{AIP/123-QED}

    \title{Silver Electrodeposition from Ag/AgCl Electrodes: Implications for Nanoscience}


\author{Chuhongxu Chen}
\email{chuhongxu.chen@postgrad.manchester.ac.uk}
\affiliation{Department of Physics and Astronomy, University of Manchester, Manchester, M13 9PL, UK}

\author{Ziwei Wang}
\affiliation{Department of Physics and Astronomy, University of Manchester, Manchester, M13 9PL, UK}

\author{Guilin Chen}
\affiliation{Department of Physics and Astronomy, University of Manchester, Manchester, M13 9PL, UK}

\author{Zhijia Zhang}
\affiliation{Department of Physics and Astronomy, University of Manchester, Manchester, M13 9PL, UK}

\author{Zakhar Bedran}
\affiliation{Department of Physics and Astronomy, University of Manchester, Manchester, M13 9PL, UK}

\author{Stephen Tipper}
\affiliation{Department of Physics and Astronomy, University of Manchester, Manchester, M13 9PL, UK}

\author{Pablo Díaz-Núñez}
\affiliation{Department of Physics and Astronomy, University of Manchester, Manchester, M13 9PL, UK}

\author{Ivan Timokhin}
\affiliation{Department of Physics and Astronomy, University of Manchester, Manchester, M13 9PL, UK}

\author{Artem Mishchenko} 
\email{artem.mishchenko@manchester.ac.uk}
\affiliation{Department of Physics and Astronomy, University of Manchester, Manchester, M13 9PL, UK}
\affiliation{National Graphene Institute, University of Manchester, Manchester, M13 9PL, UK}

\author{Qian Yang} 
\email{qian.yang@manchester.ac.uk}
\affiliation{Department of Physics and Astronomy, University of Manchester, Manchester, M13 9PL, UK}
\affiliation{National Graphene Institute, University of Manchester, Manchester, M13 9PL, UK}


\begin{abstract}
With the advancement of nanoscience, silver/silver chloride (Ag/AgCl) electrodes have become widely utilised in microscale and nanoscale fluidic experiments, because of their stability.  However, our findings reveal that the dissolution of AgCl from the electrode in \ch{Cl-}-rich solutions can lead to significant silver contamination, through the formation of silver complexes, \ch{[AgCl_{n+1}]^{n-}}. 
We demonstrate the electrodeposition of silver particles on graphene in KCl aqueous solution, with AgCl dissolution from the electrode as the sole source of silver. This unexpected electrodeposition process offers a more plausible interpretation of the recently reported ``ionic flow-induced current in graphene''. That is, the measured electronic current in graphene is due to the electrodeposition of silver, challenging the previously claimed ``ionic Coulomb drag''. More caution is called for when using Ag/AgCl electrodes in microfluidic, and especially nanofluidic systems, because AgCl dissolution should not be neglected.
\end{abstract}
\maketitle

\section{\label{sec:level1}Introduction}
The Ag/AgCl electrode is one of the most reliable reference electrodes used in electrochemistry, known for its stable electrode potential, low polarisation, and reversible redox reactions\cite{intro}. In recent years, the Ag/AgCl electrode has repeatedly demonstrated its versatility in the advancing fields of two-dimensional (2D) membranes, biosequencing, nanofluidics, and their intersections. The research highlights encompass graphene\cite{2d1, 2d2, 2d3}, nanopores\cite{nanopore1, nanopore2, nanopore3, nanopore4, nanopore5, nanopore6, nanopore7}, nanochannels\cite{nanochannel1, nanochannel2}, biosensors\cite{dna1, dna2, dna3, brainsensor}, and ionic Coulomb drag\cite{ioncoulombdrag, ioncoulomddrag2}. The Ag/AgCl electrode is considered optimal in these cases because of its local solid-to-solid electrode reactions and low solubility product constant ($K_{sp}$)\cite{Ksp}, and the contamination introduced to the experimental system was believed to be minimal. Therefore, the dissolution of AgCl in these systems, particularly involving \ch{Cl-} ions, has been overlooked so far in most studies.  Using an electrochemical cell consisting of Ag/AgCl and monolayer graphene electrodes in the KCl aqueous solution, we report the direct observation of silver electrodeposition on graphene. As the sole source of silver in this system, the Ag/AgCl electrode induced significant and unexpected silver contamination. The electrodeposition of silver becomes even more important during our investigation of the ``ionic Coulomb drag" phenomenon, where the electronic current in graphene is induced as a result of the silver electrodeposition, rather than the purported Coulomb interactions. Our findings serve as a reminder for future studies involving small-volume, ion-sensitive systems with exposed Ag/AgCl electrodes, as unintentional silver contamination could interfere with the intended experiments.

\section{\label{sec:level1}Results and Discussion}

\begin{figure}
\includegraphics[scale=0.75]{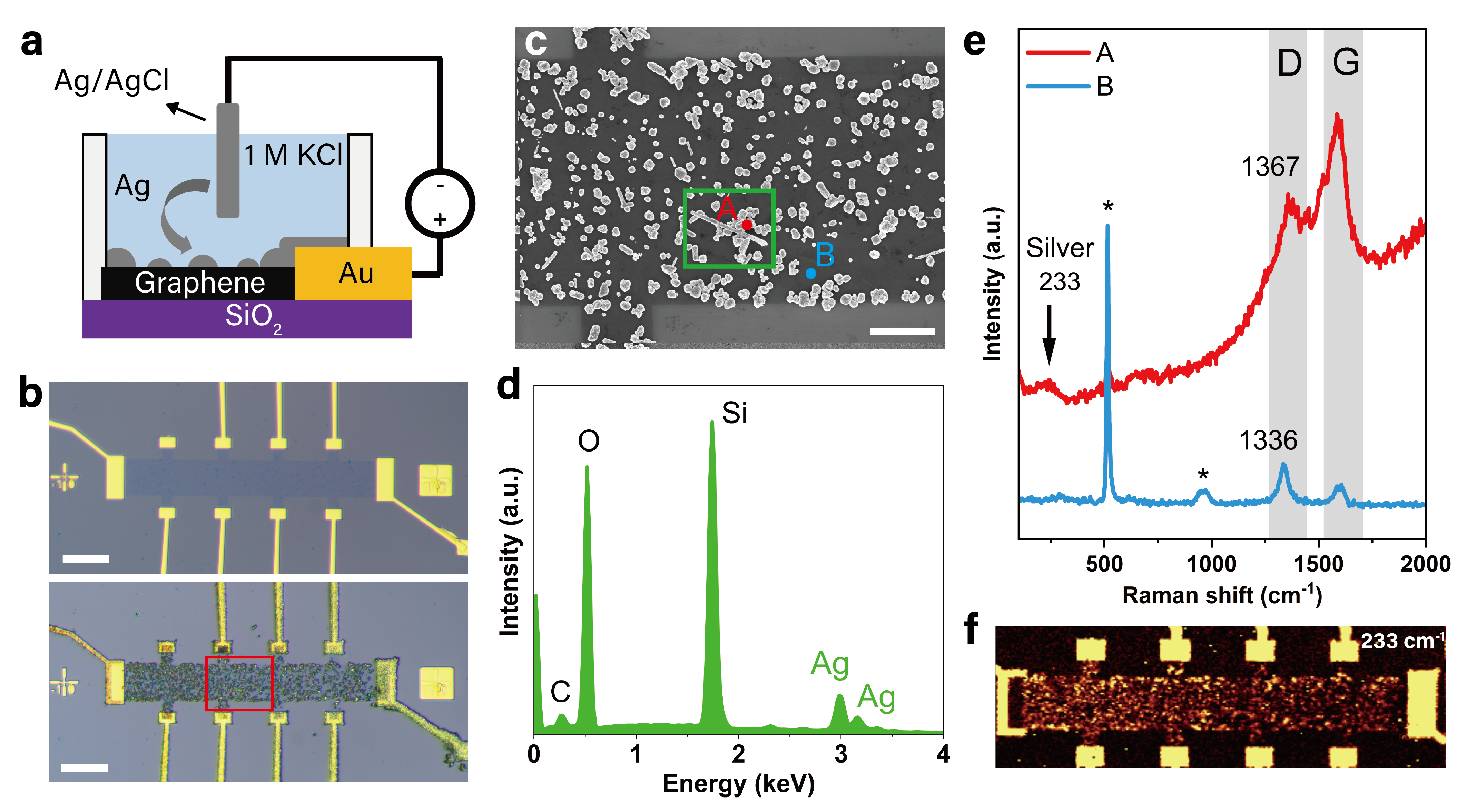}
\caption{\label{fig1} \textbf{Characterization of the device after electrodeposition.} \textbf{a}, Schematic of the electrodeposition setup. An exposed Ag/AgCl electrode works as the counter electrode, and a monolayer CVD graphene flake works as the working electrode. \textbf{b}, Optical images of the graphene device before (top) and after (bottom) the electrodeposition. Gold contacts were exposed to the solution during the experiment. Scale bar 25 $\mu$m. \textbf{c}, SEM image of the area marked by the red square in \textbf{b}. Scale bar 5 $\mu$m. \textbf{d}, EDX spectrum collected among the area marked by the green square in \textbf{c}. \textbf{e}, Raman spectra of graphene with silver particles (A) and pristine graphene (B), collected on the spots marked by A and B in \textbf{c}. The peaks marked by the asterisk are from SiO$_2$/Si substrate. \textbf{f}, Raman intensity map of the peak at 233 cm$^{-1}$ which shows the distribution of silver across the sample.}
\end{figure}

Using a two-electrode electrochemical cell (Fig. \ref{fig1}\textbf{a}), we demonstrate the electrodeposition of silver on monolayer graphene, characterised by optical microscopy, energy-dispersive X-ray spectroscopy (EDX), and Raman spectroscopy. Figure \ref{fig1}\textbf{b} shows the graphene device, shaped in a Hall bar geometry similar to a typical GrFET sensor commonly used in other studies. A lab-made Ag/AgCl electrode (see Methods) was used to electrodeposit silver in a 1M KCl solution with a constant bias voltage of 200 mV for 4 hours. During the process, the gold contacts were exposed to the aqueous solution, with graphene connected to the circuit via the gold contact on the left. Figure \ref{fig1}\textbf{b} shows the optical micrograph of the same device before and after electrodeposition (see also Fig. S5a-b, the optical images of an exfoliated graphene device before and after silver electrodeposition). Black particles are seen on both graphene and gold contacts in the latter, showing significant deposition. Figure \ref{fig1}\textbf{c} shows the scanning electron microscopy (SEM) image of the region marked by the red square of Fig. \ref{fig1}\textbf{b}. EDX analysis, shown in Fig.\ref{fig1}\textbf{d} (mapping see supplementary Fig. S1), demonstrates profound Ag peaks, confirming that the deposited particles contain silver. Silver is known to be thermodynamically unstable and readily oxidises under ambient conditions. For silver nanoparticles, this oxidation occurs even more rapidly\cite{Agunstable}. Therefore, the deposited particles during the measurement are highly likely to have already undergone oxidation, forming a mixture of silver compounds. Additionally, to rule out potential defects in our lab-made electrodes, a commercially available Ag/AgCl electrode was also used in later experiments, and a similar electrodeposition was observed, see Fig. S4.

We also conducted Raman spectroscopy to characterise the silver-deposited graphene. Figure \ref{fig1}\textbf{e} shows the Raman spectra collected over two positions on the device, A and B, with and without silver particles, as marked in Fig. \ref{fig1}\textbf{c}. The silver particle greatly amplifies the Raman signal of graphene, causing a more significant background, stronger G and D bands, and a slight shift in the D band from 1336 cm$^{-1}$ to 1367 cm$^{-1}$. This amplification can be explained by the localised surface plasmon resonance (LSPR) of metal particles, which is also the fundamental working principle of the surface enhanced Raman scattering (SERS) technique\cite{SERS}. To map the landscape of deposited silver, we highlight a band at 233 cm$^{-1}$, which has been observed in many silver compounds\cite{Agplasmon, 233a, 233b, 233c, 233d, 233e}. Although studies usually associate this band with the Ag-Cl\cite{233b, 233c, 233e} or Ag-O\cite{233c, 233d} vibration, our observation aligns more with the Ag$^0$ plasmonic resonance\cite{Agplasmon}, with more details shown in the supplementary information. Figure \ref{fig1}\textbf{f} presents an integrated intensity map of the Raman band centred at 233 cm$^{-1}$ to reflect the silver distribution across the graphene sample after electrodeposition. It reveals that, unlike the uniform silver layer deposited on gold, the silver deposited on graphene aggregates into discrete small islands, consistent with the SEM images in Figure \ref{fig1}\textbf{c}. Our observations demonstrate the extent to which an Ag/AgCl electrode can contaminate an experimental system by introducing undesired silver-containing compounds to the solution and surfaces.

The electrodeposition of silver on graphene requires the dissolution of silver ions into the solution. To understand the excessive dissolution of AgCl in our system, we note that the common ion effect assumes that AgCl dissolves only to its constituent ions \ch{Ag+} and \ch{Cl-} with an equilibrium of $K_{sp} = [\ch{Ag+}][\ch{Cl-}]$, where $K_{sp} = 1.8 \times 10^{-10}$ is the solubility product of AgCl at 25$^{\circ}$C\cite{Ksp} and square brackets represent the concentration in mol/l. Using this formula, \ch{Ag+} introduced by AgCl dissolution is approximately $1.8 \times 10^{-10}$ M in 1 M KCl solution, which can be safely disregarded in most systems. However, the dissolution of AgCl in \ch{Cl-}-rich solutions exceeding what is expected from the common ion effect, has been known since the last century in the study of AgCl electrode degradation\cite{agclproblem}. Instead of forming \ch{Ag+} and \ch{Cl-} directly, AgCl forms \ch{[AgCl_{n+1}]^{n-}} complexes in the presence of \ch{Cl-}, increasing its solubility. At room temperature, AgCl dissolves in pure water at a concentration of $1.3 \times 10^{-5}$ M, which is already five orders of magnitude higher than expected under the common ion effect assumption. In solutions with increasing \ch{Cl-} concentration, the solubility of AgCl further increases by orders of magnitude, reaching up to $2.4 \times 10^{-3}$ M in a 3 M KCl solution\cite{solubility2} without a change in pH.

\begin{figure}
\includegraphics[scale=0.75]{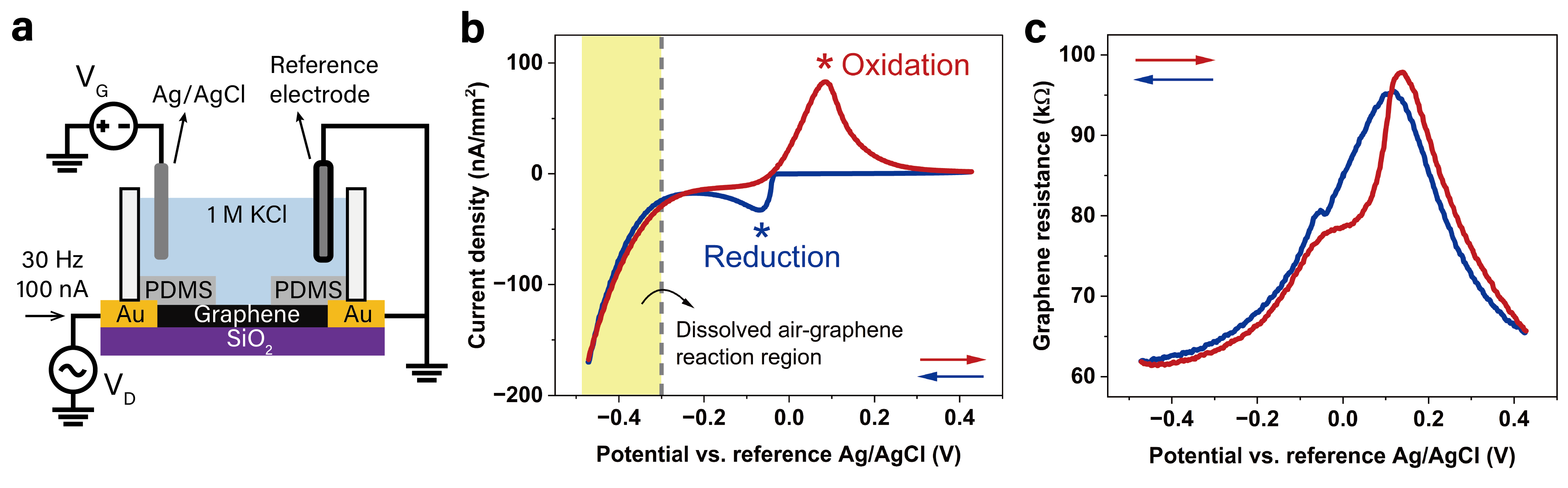}
\caption{\label{fig2} \textbf {Electrodeposition process using three-electrode electrochemical cell.} \textbf{a}, Schematic of the experimental setup for gating graphene and performing cyclic voltammetry. The setup consists of an Ag/AgCl counter electrode, a reference electrode and a graphene working electrode, forming a three-electrode electrochemical cell. Graphene resistance is monitored by applying a drain current of 100 nA at 30 Hz. V$_G$ is the gate voltage, and V$_D$ is the drain voltage. \textbf{b}, Cyclic voltammogram and \textbf{c}, graphene gating effect plotted against the gate potential (graphene potential vs. reference electrode). Red and blue arrows indicate the direction of the potential sweep. Asterisks in \textbf{b} mark the oxidation and reduction peaks. The shaded yellow region highlights the reduction reaction of dissolved air in the solution at the graphene surface.
}
\end{figure}

We also conducted cyclic voltammetry (CV) to investigate the electrodeposition of silver on graphene using a three-electrode electrochemical cell, measured in 1 M KCl solution at a scan rate of 200 mV/min, as shown in Fig. \ref{fig2}\textbf{a}. In this setup, a large, millimetre-sized CVD graphene served as the working electrode. A patterned PDMS film was applied to protect the gold contacts from the electrolyte, ensuring that electrochemical reactions occurred exclusively on graphene. The exposed graphene area was approximately 6 mm$^2$. The reference electrode was an Ag/AgCl wire immersed in an isolated 3.4 M KCl solution, connected to the rest of the system through a leak-free junction manufactured by Innovative Instruments, Inc. This specialised junction prevents the passage of ions, effectively eliminating cross-contamination. A Keithley 2614 sourcemeter was used to apply a voltage between graphene and the exposed Ag/AgCl counter electrode, and the potential difference between graphene and the reference electrode was monitored using a Keithley 2182A nanovoltmeter. This setup not only allows us to monitor the electrochemical process but also allows for the simultaneous gating of graphene. To evaluate the gate effect, a drain current of 100 nA at 30 Hz was applied, and the changes in graphene resistance were measured using an SR860 lock-in amplifier.

Figure \ref{fig2}\textbf{b} represents the cyclic voltammogram of the electrodeposition processes. The voltammogram shows one reduction peak and one oxidation peak, corresponding to the redox reaction of silver. These features indicate reaction dynamics similar to those previously reported for silver electrodeposition on HOPG\cite{HOPG}, ITO-coated glass\cite{glass}, and platinum\cite{plat} using \ch{AgNO3} solutions. In our system, the absence of \ch{NO3-} means that silver can only be oxidised to AgCl, a compound of very low solubility compared to \ch{AgNO3}.  Since silver ions in the solution originate primarily from the dissolution of the exposed Ag/AgCl electrode, any silver lost in the solution due to electrodeposition is rapidly replenished by the continuous dissolution of the Ag/AgCl electrode in the KCl solution. During this process, a sheer increase in the cathodic current is observed at negative potentials below -0.3 V (marked as the shaded yellow region). This can be partially attributed to the electrodeposition current. However, the reduction of oxygen from the dissolved air in an aqueous solution on the graphene surface could also contribute largely to this increase, as previously reported\cite{leak}. This reaction effectively enlarges the existing defects in graphene, aligning well with our observations, as CVD graphene samples consistently lose their integrity after prolonged CV scans (Fig. S3\textbf{a}). The applied potential not only drives the electrochemical reaction but also gates graphene through the electrical double layer. This gating effect is demonstrated in Fig. \ref{fig2}\textbf{c}, where the graphene resistance reaches a maximum at around 0.11 V, corresponding to the charge neutrality point of graphene. However, noticeable shoulders in the curves suggest strong inhomogeneity of the sample, as a result of the possible damage to graphene during continuous measurements. In the context of gating graphene, the current in Fig. \ref{fig2}\textbf{b} can also be interpreted as the gate ``leakage current'', which should ideally be zero. Therefore, in liquid gating systems, it is recommended to employ a lock-in technique to minimise the interference of the DC leakage current with the drain current measurements.

\begin{figure}
\includegraphics[scale=0.7]{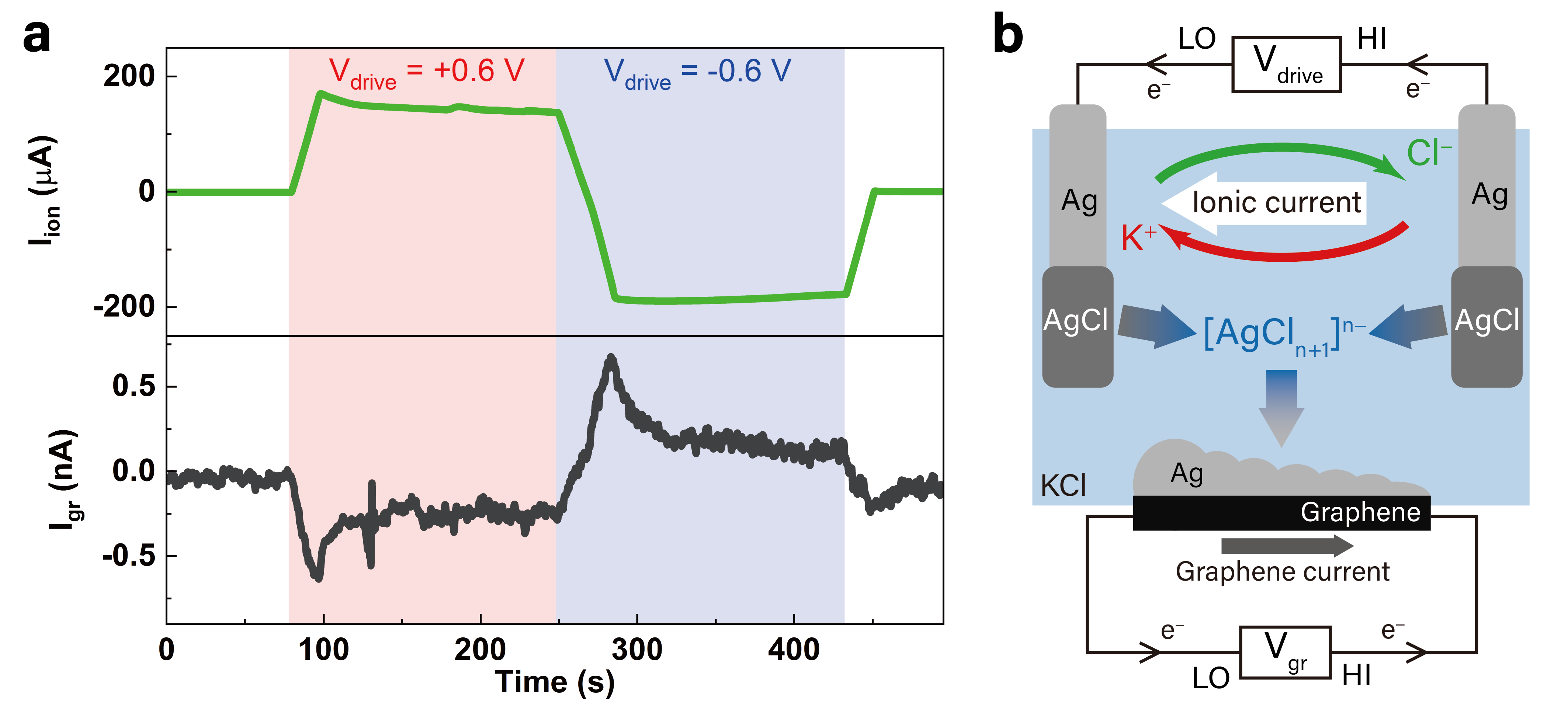}
\caption{\label{fig3} \textbf{Electron current in graphene induced by silver electrodeposition.} \textbf{a}, Ionic current, I$_{ion}$ (upper panel), generated between two Ag/AgCl electrodes with $\pm$0.6 V driving voltage and a drag-like electronic current, I$_{gr}$ (lower panel), induced in graphene. Regions coloured in red and blue mark the driving voltage with opposite polarity. \textbf{b}, Illustration of the current generation in graphene via electrodeposition of silver in our system. V$_{drive}$ and V$_{gr}$ are the drive voltage of the ion flow in the KCl solution and the voltage induced in graphene respectively. The electrochemical reactions on two Ag/AgCl electrodes are \ch{AgCl(s) + e- --> Ag(s) + Cl-(aq)} and \ch{Ag(s) + Cl-(aq) --> AgCl(s) + e-}. The electrochemistry in the system can generate a graphene current with a reversed sign to the ionic current based on the electrodeposition of silver (not Coulomb drag). Arrows in green and red indicate the flow of Cl$^-$ and K$^+$ ions. Arrows in grey/blue represent the formation and deposition of [AgCl$_{n+1}$]$^{n-}$. 
}
\end{figure}

We have shown that in a system with concentrated \ch{Cl-}, the dissolution of AgCl rapidly introduces silver-containing ions into the solution, making silver contamination a significant issue. Silver contaminants may further interfere with the intended measurements, which, if overlooked, may lead to misinterpretation of experimental results. To highlight its impact, we carried out experiments in which ionic flow along the graphene surface induces an electronic current in graphene, which was previously attributed to the ``ionic Coulomb drag\cite{ioncoulombdrag}.'' Ionic Coulomb drag refers to the transfer of momentum from moving ions in a solution, to electrons in a nearby solid, such as graphene, through Coulomb interactions. Several recent studies\cite{ioncoulombdrag, ioncoulombdrag3, ioncoulombdrag4} have reported the observation of this effect in microfluidic systems, underlining its potential for energy harvesting and sensing applications.

To investigate this phenomenon, different experimental configurations were used. Figure \ref{fig3}\textbf{b} shows one measurement configuration. In this setup, a millimetre-sized CVD graphene was positioned between two Ag/AgCl electrodes in a liquid channel filled with 1 M KCl solution. An ionic flow was generated by applying a voltage bias across a pair of Ag/AgCl electrodes. At each electrode, the \ch{Ag(s) + Cl- <-> AgCl(s) + e-} reaction locally generates or consumes silver ions (depending on bias polarity), causing \ch{Cl-} to migrate between two electrodes, accompanied by a corresponding \ch{K+} flow\cite{agclreaction1, agclreaction2}. In our experiments, $\pm$0.6 V voltage was applied between two Ag/AgCl electrodes using one channel of a Keithley 2614 sourcemeter, and the current generated in graphene was measured using the second channel of the sourcemeter.

This measurement setup closely followed the methodology of an earlier study\cite{ioncoulombdrag}, and indeed, we observed similar signals. The upper panel of Fig. \ref{fig3}\textbf{a} shows the ionic current generated in the solution, while the lower panel shows the electronic current induced in graphene in response to the ionic current. With a $\mu$A-level ionic current, a sign-reversed nA-level electronic current is observed in graphene. The observed characteristics, particularly the sign-reversed electronic current, are consistent with the previously proposed ``ionic Coulomb drag" explanation. However, after repeating the same experiments on dozens of similar graphene devices, we observed significant variations in the measured results, both in magnitude and sign-reversal behaviour. In fact, in half of the tested devices, the measured electronic current became sign-aligned with the ionic current (Fig. S5 of the supplementary information). Such randomness makes the ``ionic Coulomb drag" untenable and an alternative interpretation is needed. 

In the experiment, although the Ag/AgCl electrodes and graphene are electrically isolated, they remain coupled through the electrodeposition process, allowing ions and electrons to interchange. Importantly, the sourcemeter connected to graphene can act as an electron source or sink even without applying a voltage, supplying electrons necessary for silver electrodeposition. Hence, we propose that the electronic current detected in graphene is solely due to silver deposition rather than Coulomb drag! See Fig. S3b for the optical micrograph of the graphene device after measurements, with profound silver-containing particles deposited on graphene. Fig. \ref{fig3}\textbf{b} illustrates this mechanism: AgCl dissolves in the KCl solution, forming silver complexes \ch{[AgCl_{n+1}]^{n-}}. Under an applied voltage between the Ag/AgCl electrodes, generating an electronic current.  If the deposition were perfectly even (which is unlikely), the currents would cancel out, and no net current would appear. In reality, the deposition is always uneven, so we detect a current. The strength and direction of this current depend on how much silver is deposited and where it starts. Variations in current magnitude and sign arise from differences in deposition extent and location on graphene, explaining the observed inconsistencies across multiple devices (cf. \ref{fig3}\textbf{b} and Fig. S5 in the supplementary information). Thus, the generation of electronic current in graphene depends exclusively on silver redox reactions, independent of direct ionic interactions with graphene.

In summary, we demonstrate that the dissolution of AgCl in \ch{Cl-}-rich aqueous systems can introduce serious contamination, which, if overlooked, could lead to misinterpretation of experimental results in such systems. Ag/AgCl electrodes, despite their frequent usage in nanoscience, could cause significant complications to the system. Care should be taken when applying Ag/AgCl electrodes in ion-sensitive environments, especially with high \ch{Cl-} concentrations. Ag/AgCl electrodes should be isolated appropriately using low-leak or leak-free junctions, even when used only as reference electrodes. 

\section{\label{sec:level1}Methods}
The CVD graphene grown on copper (Cu) used in this study was purchased from Graphenea. For electronic device fabrication, graphene was transferred onto \ch{SiO_{2} (290 nm)/Si} substrate by a poly-methyl methacrylate (PMMA)-assisted method, described elsewhere\cite{CuEtching}. In brief, the Cu substrate was etched overnight in a 20 g/L ammonium persulfate solution. Oxygen plasma dry etching was then employed to shape the transferred graphene. For small devices (less than 200 $\mu$m), a PMMA mask was patterned using standard electron beam lithography, while large devices on the millimetre scale were shaped using a plastic mask. Finally, Cr (3 nm)/Au (40 nm) electrodes were deposited via electron beam metal evaporation. All fabrication processes were performed in a cleanroom environment. 

The Ag/AgCl electrodes were either lab-made or purchased from World Precision Instruments. For the fabrication of lab-made electrodes, silver wire (0.8 mm diameter, 99.99\% purity) supplied by GoodFellow was used. Prior to electroplating, the silver wires were polished with sandpaper and cleaned by ultrasonication in acetone and deionised water. The cleaned silver wires were then immersed in a 1 M KCl solution, and a constant current of 1 mA was applied between them until a stable potential was reached, which required approximately 30 minutes. During this process, AgCl was electroplated onto the cathode.

Raman spectroscopy was conducted using WITec Alpha 300R Raman imaging microscope (Oxford Instruments) with a 532 nm excitation wavelength and a $\times$50 magnification lens (NA = 0.75, spot size = 364 nm). System control and data analysis were performed using the WITec suite 6.1 software. Raman mapping was performed over a 160 $\mu$m $\times$ 50 $\mu$m area with 0.5 $\mu$m step size, using laser power of 1.5 mW, 1.7 s integration time, and 1 accumulation. The map shown in Fig.\ref{fig1}\textbf{d} was obtained by applying a sum filter to the peak at 233 cm$^{-1}$ in Fig.\ref{fig1}\textbf{e} with a width of 150 cm$^{-1}$. The background was subtracted with a slope defined by 4 pixels on both sides of the filter range. To enhance clarity, the maximum intensity was capped at 700 CCD counts, as the Raman signal from silver deposited on gold was
significantly stronger than that from silver deposited on graphene.

Scanning electron microscopy (SEM) imaging was carried out using a Zeiss Ultra Field Emission SEM equipped with Oxford Instruments X-Max detector for EDX measurements. SEM images were acquired using an in-lens electron detector and 30 $\mu$m aperture. EDX measurements were performed at an acceleration voltage of 7.05 kV to analyse the Ag L$\alpha$ electron (energy: 2.983 keV). To ensure accuracy, the EDX spectrum was averaged over ten acquisitions.

\bibliography{aipsamp}

\newpage

\renewcommand{\figurename}{Fig.}
\renewcommand*{\thefigure}{S\arabic{figure}}
\setcounter{figure}{0}

\section*{Supporting Information for Silver Electrodeposition from Ag/AgCl Electrodes: Implications for Nanoscience}
\subsubsection{EDX mapping of the silver-deposited graphene}

\begin{figure}
\includegraphics[scale=0.14]{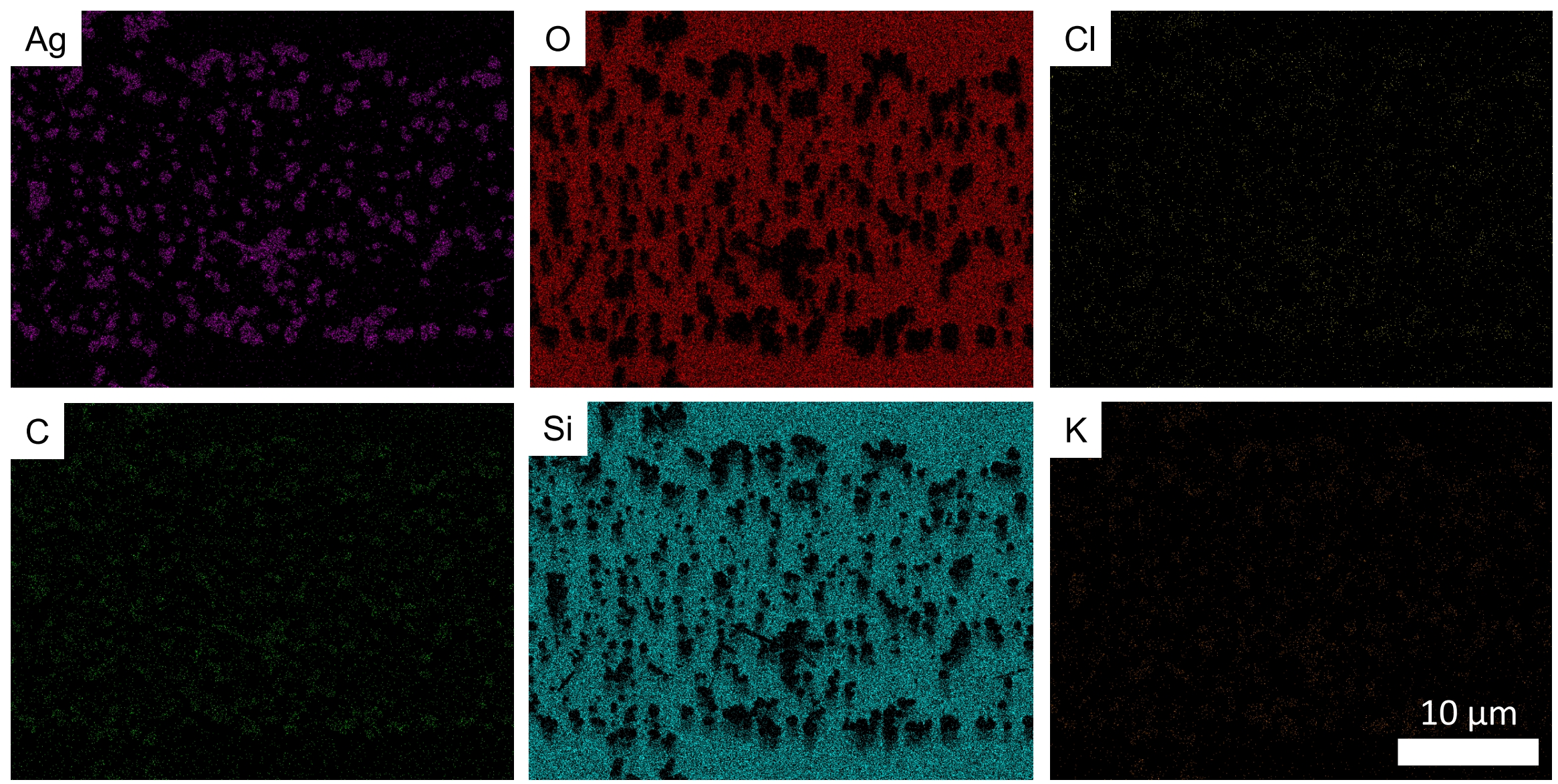}
\caption{\label{figs1} Energy dispersive X-ray (EDX) mapping of composites, measured together with Fig. 1\textbf{d} of the main text.}
\end{figure}

Figure \ref{figs1} presents the energy-dispersive X-ray (EDX) colour mapping for the sample shown in Fig. 1\textbf{c} of the main text, with elemental distributions plotted for Ag, C, O, Si, K, and Cl. The colour maps indicate that the particles on graphene primarily consist of metallic silver, as the Cl and O signals for AgCl and \ch{Ag2O} are insignificant in these regions. The Si and O signals predominantly originate from the \ch{SiO2} substrate, the C signal corresponds to graphene, and the K and Cl signals can be attributed to residual KCl from the solution. EDX mapping provides a qualitative distribution of elements. However, because of the peak overlap among the analysed elements (peaks used for analysis: Ag: 2.984, 3.151, 2.806, 2.634, 0.312 keV; Cl: 2.622, 2.816, 0.184 keV; K: 3.314, 3.590, 0.262 keV), it is hard to quantitatively confirm the exact components of the silver-containing compounds deposited on graphene. 

\subsubsection{Raman spectroscopy of Ag and AgCl}



\begin{figure}
\includegraphics[scale=0.1]{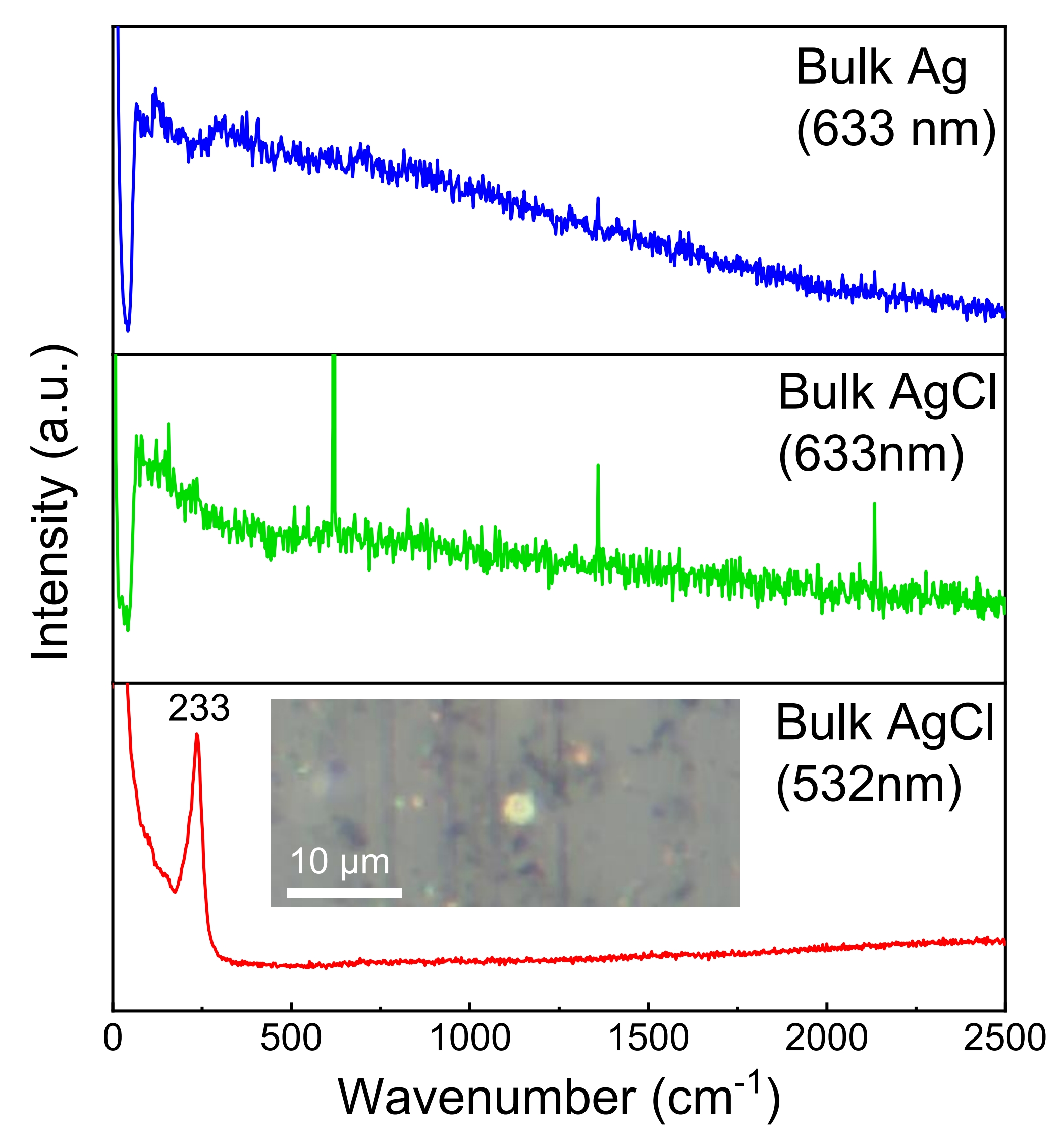}
\caption{\label{figs2} \textbf{Raman spectroscopy of bulk Ag and AgCl.} A red laser (633 nm) and a green laser (532 nm) were used for the AgCl sample, where the green laser reduced AgCl to Ag, resulting in the emergence of a peak at 233 cm$^{-1}$. Inset: Optical micrograph of the AgCl sample after exposure to the green laser.}
\end{figure}

We conducted Raman spectroscopy for pure Ag and pure AgCl samples to reveal the origin of the 233 cm$^{-1}$ band in the Raman spectra of silver-deposited graphene, shown in Fig. 1(e) of the main text. Both Ag and AgCl samples were scratched using a glass slide to remove potential surface contamination and to expose fresh surface before measurements. As shown in Fig. \ref{figs2}, no significant Raman signatures were observed for either Ag or AgCl under a 633 nm red laser. This is expected, as metals like Ag do not exhibit Raman signals, and first-order Raman scattering is forbidden in AgCl due to its Oh symmetry\cite{AgClnopeak}. However, the 233 cm$^{-1}$ band emerges in the AgCl Raman spectra taken with a 532 nm laser. We also observed that 532 nm laser exposure can transform the original grey-coloured AgCl into a shiny appearance, as shown in the inset. We attribute this to the light-sensitive nature of AgCl, which, upon irradiation with a high-energy green laser, is reduced to Ag. Therefore, the 233 cm$^{-1}$ band is associated to the localized surface plasmon resonance of Ag, as the AgCl reduction can generate a porous Ag nanostructure\cite{AgClreduction}.

\subsubsection{Silver electrodeposition on CVD and mechanically exfoliated monolayer graphene.}

\begin{figure}
\includegraphics[scale=0.11]{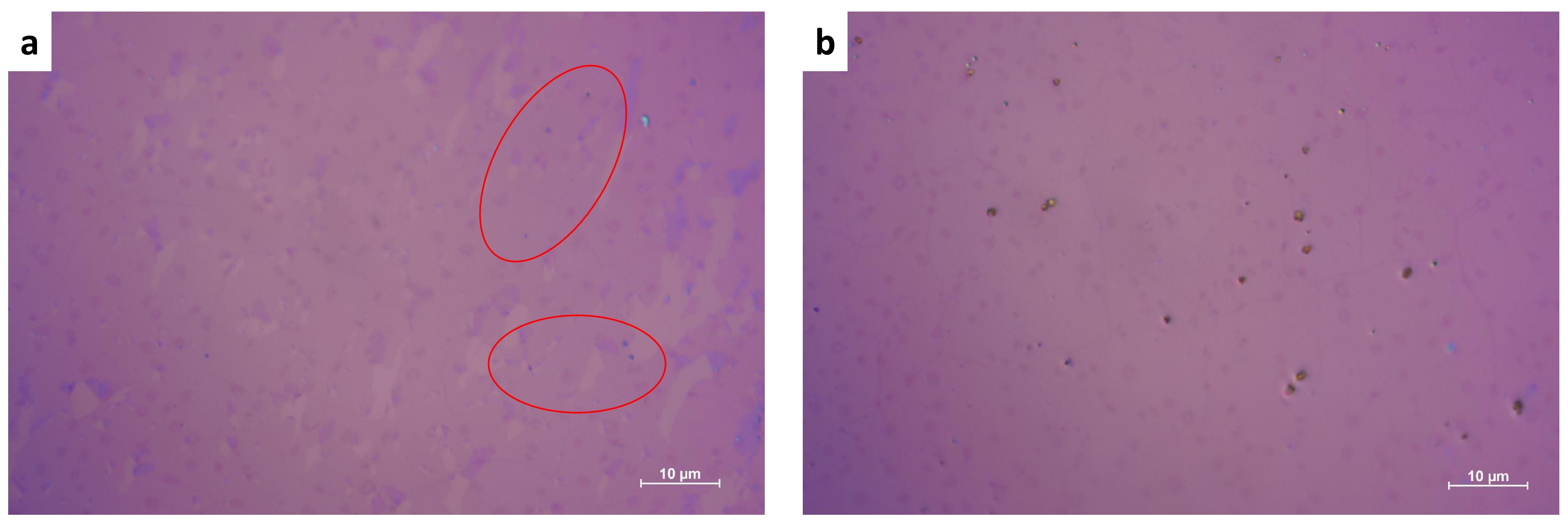}
\caption{\label{figs4} \textbf{Optical micrographs of two large CVD graphene devices used in} \textbf{a}, Fig. 2 and \textbf{b}, Fig. 3 of the main text, taken after measurements. Nano-sized silver-containing particles are highlighted with circles in \textbf{a}.  }
\end{figure}

With hindsight, it is now easy for us to spot silver electrodeposition and to attribute the observed electronic current in graphene to the electrodeposition of silver-containing particles. However, we first noticed this issue only after measuring more than 10 devices. Here we show in Fig. \ref{figs4}\textbf{a} and \textbf{b} the graphene flakes used in Fig. 2 and Fig. 3 of the main text after measurements, respectively. \ref{figs4}\textbf{a} shows the optical images of two devices after measurements using the commercially available Ag/AgCl electrode. After cyclic voltammetry measurements, nanometre-sized silver-containing particles are electrodeposited on the graphene surface (marked by the circled area in Fig. \ref{figs4}\textbf{a}). Because of much milder deposition conditions - shorter time and lower voltage, less deposition is seen, which may be hard to identify and can be easily overlooked. The raptures in the graphene were formed during the measurement and are likely caused by the reaction of dissolved air at the graphene surface, as has been previously reported\cite{leak}. The device in Figure \ref{figs4}\textbf{b} was used to investigate the ``ionic Coulomb drag'', where we measured the graphene current under a constant ionic current for a relatively long period. This experiment represents a more stable electrodeposition process, leading to an increased amount of electrodeposition on the graphene surface. This image supports our statement that the silver electrodeposition current contributes to the graphene current measured in Fig.3\textbf{b} of the main text.

\begin{figure}
\includegraphics[scale=0.15]{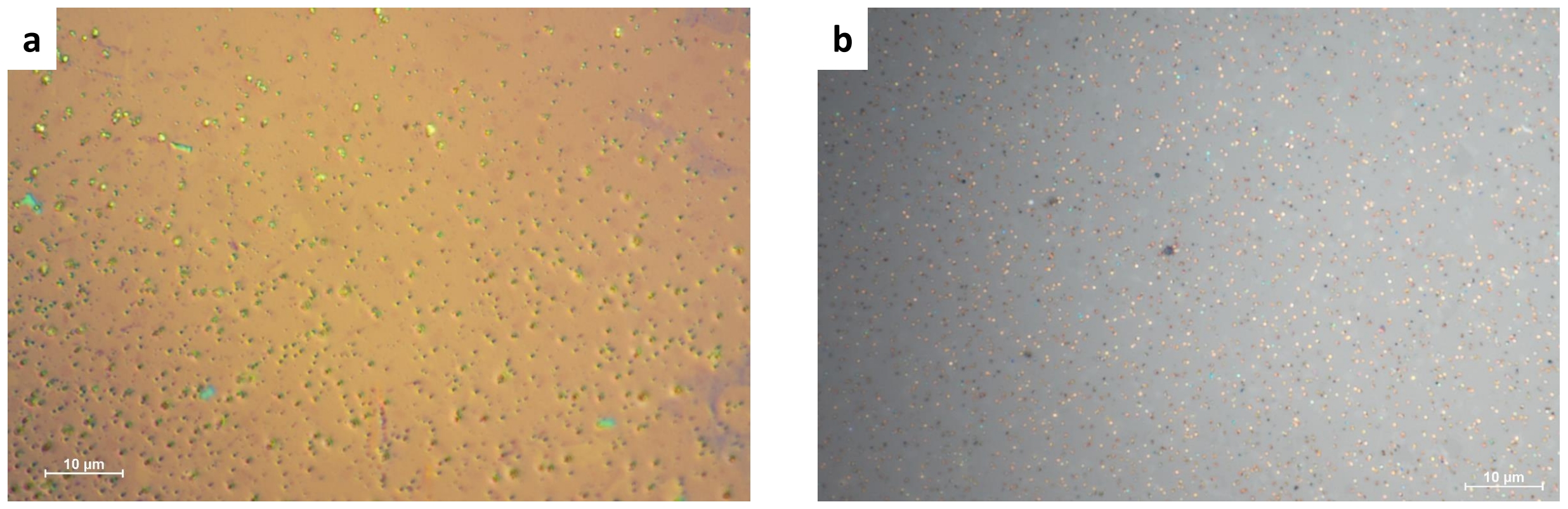}
\caption{\label{commercialaggraphene} \textbf{Optical micrographs of two large CVD graphene devices deposited with silver using the commercial Ag/AgCl electrode.} \textbf{b}, Silver-deposited graphene on a Si/\ch{SiO2} substrate. \textbf{a}, Silver-deposited graphene on a quartz substrate.}
\end{figure}

To demonstrate that the issue with Ag/AgCl electrodes is not limited to our lab-made electrodes, we show here two millimetre-sized CVD graphene deposited with silver using the commercial Ag/AgCl electrode in pure KCl solutions. Figure \ref{commercialaggraphene}\textbf{a} shows a graphene device after several hours of CV measurement, while Fig. \ref{commercialaggraphene}\textbf{b} depicts another graphene device following silver deposition under a 0.25 V constant bias voltage for one hour. Both samples exhibit a substantial presence of nanosized silver particles, indicating strong electrodeposition. The device in panel \textbf{a} was fabricated on a Si/\ch{SiO2} substrate, while the device in panel \textbf{b} was fabricated on a transparent quartz substrate, resulting in colour differences in their optical images. Additionally, the silver particles in panel \textbf{b} did not undergo any CV measurement and, therefore, were never oxidized to AgCl. This explains why they appear shinier than the silver particles in panel \textbf{a}.

\begin{figure}
\includegraphics[scale=0.12]{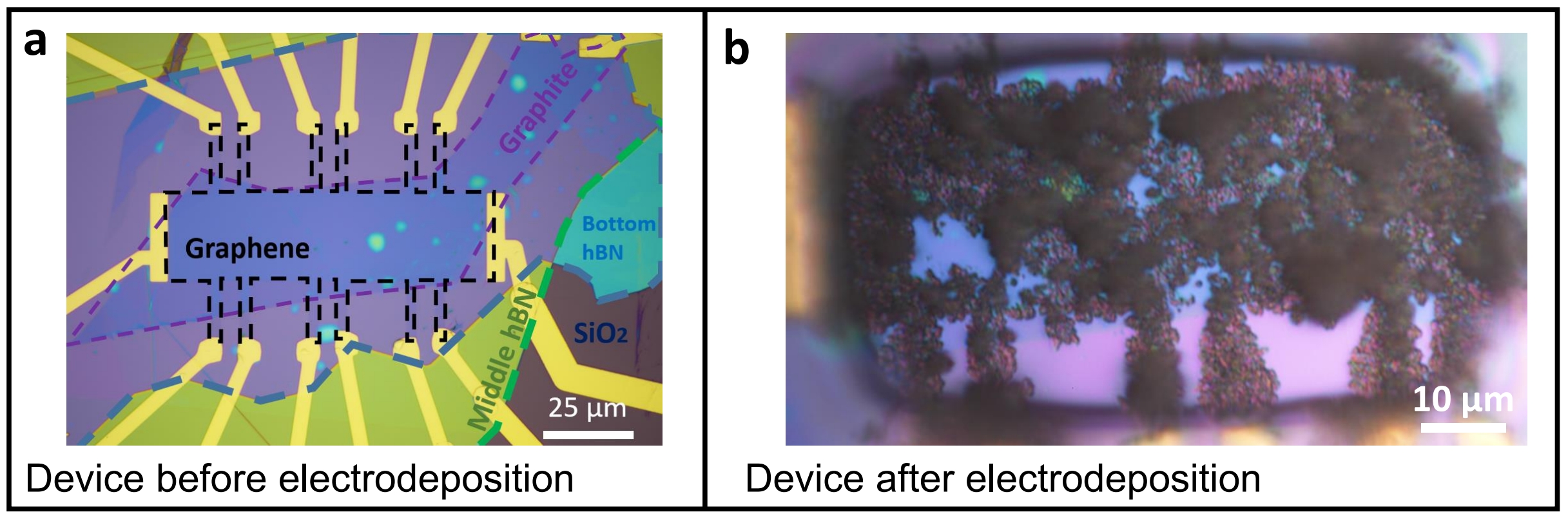}
\caption{\label{figs3}  \textbf{Electrodeposition on mechanically exfoliated graphene device.} \textbf{a}, Optical micrographs of the heterostructure device featuring a mechanically exfoliated graphene hall bar, with each layer outlined by dashed lines. \textbf{b}, Deposition of Ag-containing nanoparticles on the graphene surface after prolonged electrodeposition. The particles demonstrate inhomogeneous deposition.}
\end{figure}

The electrodeposition of silver on mechanically exfoliated graphene was also observed.  Fig. \ref{figs3}\textbf{a} shows one of such devices, which is a heterostructure fabricated using mechanically exfoliated graphene, demonstrating substantial electrodeposition after prolonged measurements. Attempts to gate graphene and to investigate the claimed ``ionic Coulomb drag'' were carried out on this device using exposed Ag/AgCl electrodes in the KCl solution. We noticed that the device was coated with a thick layer of silver composites after the measurements, as shown in Fig. \ref{figs3}\textbf{b}.

\subsubsection{Additional data of ``ionic Coulomb drag" experiments}
\begin{figure}
\includegraphics[scale=0.15]{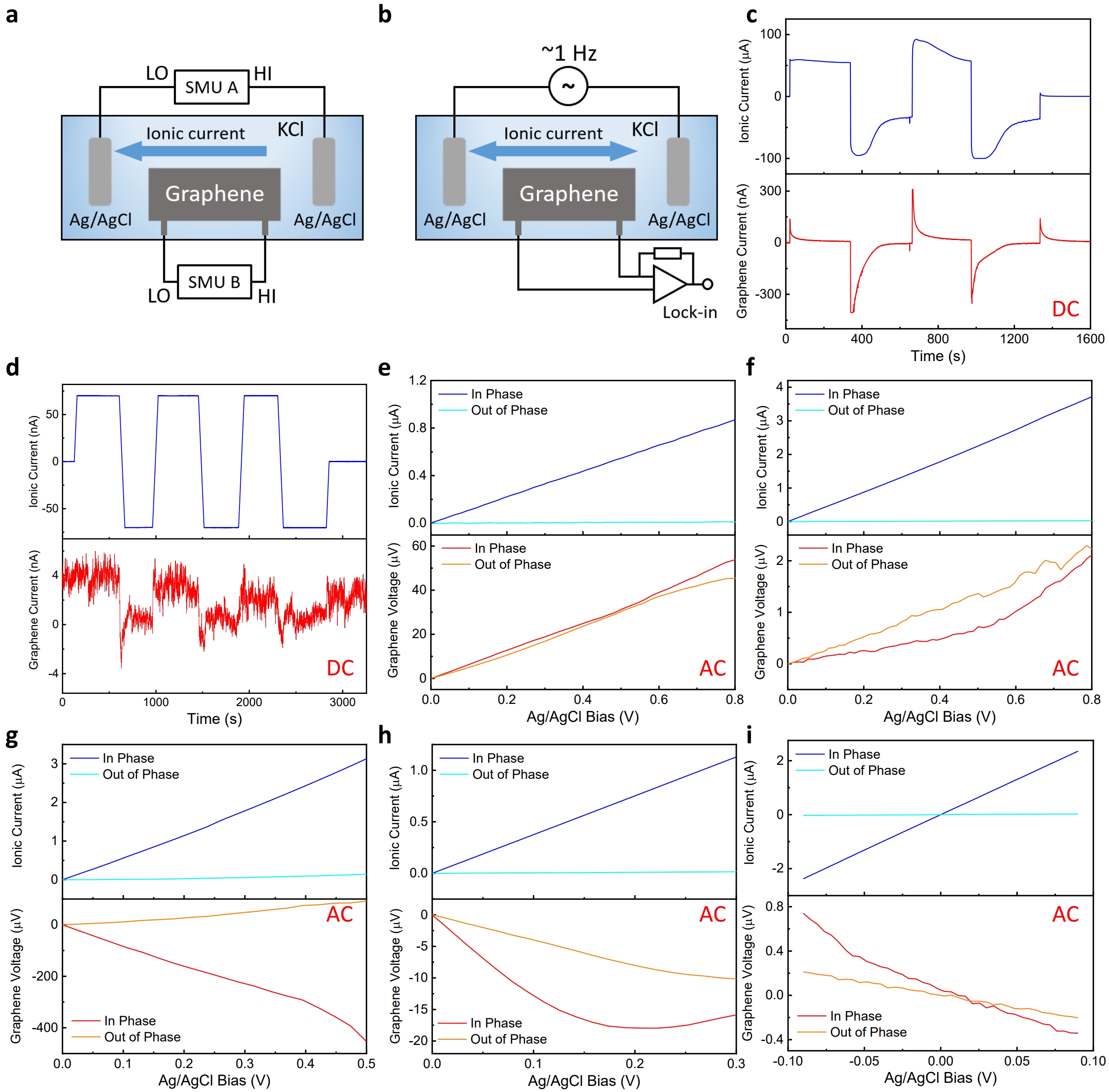}
\caption{\label{figs5} \textbf{Investigating the electronic current in graphene with ionic current.} \textbf{a}, Schematic of the DC measurement configuration, with SMU A and SMU B representing two channels of a sourcemeter. \textbf{b}, Schematic of the AC measurement configuration, where the amplifier denotes the differential voltage input of a lock-in amplifier, sharing the same oscillator (synchronized) with the AC source. \textbf{c}-\textbf{i}, Ionic currents (upper panels) and graphene currents/voltages (lower panels) measured from seven separate graphene devices. In devices \textbf{c}-\textbf{f}, the graphene signals have the same polarity as the ionic currents, while in devices \textbf{g}-\textbf{i}, they have opposite polarity.}
\end{figure}

Two measurement configurations were used to investigate the ``ionic Coulomb drag'': DC and AC, with their experimental setups shown in Fig. \ref{figs5}\textbf{a} and \textbf{b}, respectively. Details of the DC setup are provided in the main text. Briefly, one channel of Keithley 2614 sourcemeter (SMU A) drives the ionic current, while the other channel (SMU B) measures current through graphene. In the AC measurement configuration, the sourcemeter was replaced with a lock-in amplifier to minimize noise. The wave output shares the same oscillator as the measurement unit, so that the measured signal has no phase shift with the output wave. The frequencies used in the AC configuration were kept below 1 Hz to ensure that the ionic current remained entirely in phase with the driven voltage. We summarized the results from seven devices in Fig. \ref{figs5}\textbf{c}–\textbf{i}, where the upper panel displays the ionic current and the lower panel shows the graphene current. Data in \textbf{c} and \textbf{d} were measured using the DC configuration, while \textbf{e}-\textbf{i} were measured using the AC configuration. Without delving into the details of each device, we show that in three of the devices, the graphene current is sign-reversed to that of the ionic current, while in the remaining four devices, the graphene current is sign-aligned with the ionic current. Among all the devices we investigated for this phenomenon, we saw roughly half of the devices exhibiting the "sign-reversed" feature, which could lead to the ``ionic Coulomb drag'' interpretation\cite{ioncoulombdrag}. However, considering all the measurements, we believe the observed signal is caused by the silver electrodeposition process, rather than the previously proposed ionic Coulomb drag current. 

\end{document}